\documentclass[10pt]{article}

\usepackage[letterpaper, margin=1.5in]{geometry}
\usepackage{amssymb,amsmath}
\usepackage{graphicx}

\usepackage{hyperref}

\begin{document}

%\mainmatter  % start of an individual contribution

\title{Using Game Theory to Study the \\Evolution of Cultural Norms}

%\author{Soham De%
%\thanks{Dept. of Computer Science}
%\and Dana Nau%
%\thanks{Dept. of Computer Science, Institute for Systems Research, and Institute for Advanced Computer Studies}
%\and Michele Gelfand%
%\thanks{Dept. of Psychology}
%}
%\institute{University of Maryland\\
%College Park, MD 20742~ USA\\
%\url{sohamde@cs.umd.edu},~
%\url{nau@cs.umd.edu},~
%\url{mgelfand@psyc.umd.edu}
%}

\author
       {Soham De$^1$, Dana S. Nau$^{1, 2}$, and Michele J. Gelfand$^3$ \\  
       $^1$Department of Computer Science \\ $^2$Institute for Systems Research \\ $^3$Department of Psychology\\
       University of Maryland, College Park, MD, USA\\
       \texttt{\{sohamde, nau\}@cs.umd.edu, mgelfand@umd.edu}
       }


%\toctitle{Lecture Notes in Computer Science}
%\tocauthor{Authors' Instructions}
\maketitle

\begin{abstract}
We discuss how to use evolutionary game theory (EGT) as a framework for studying how cultural dynamics and structural properties can influence the evolution of norms and behaviors within a society.
We provide a brief tutorial on how EGT works, and discuss what kinds of insights it can provide.
We then describe three published studies in which we have developed EGT models
that help explain how structural and external conditions in a society affect the emergence of social norms.
%\keywords{Evolutionary game theory, cultural evolution, cultural norms}
\end{abstract}

\section{Introduction}

Understanding human behavior and modeling how cultural norms evolve in different human societies is vital for designing policies and avoiding conflicts around the world.
In this paper we discuss ways to use computational game-theoretic techniques to gain insight into why different human societies have different norms and behaviors. In particular, we will use evolutionary game-theory (EGT), which was originally developed to model the evolution of biological life forms, but also is useful for understanding the evolution of human cultures. 

EGT models typically deal with a large population of individuals that interact with each other. Different individuals have different strategies, hence their interactions lead to different
game-theoretic payoffs---which are used to represent the individuals' evolutionary fitness. 
Individuals with higher fitness (or lower fitness) will be more likely (or less likely, respectively) to reproduce, hence the strategies used by those individuals will become more prevalent (or less prevalent) in the population. In a biological model, an individual's strategy might represent its species; in a cultural model it might represent a particular kind of behavior. Thus, EGT offers a simple framework for studying the evolution of populations, without requiring the decision-theoretic ``rationality'' assumptions that are typically required in classical game theory models.

One of the limitations of EGT models is that they deliberately omit large amounts of detail.
In EGT models of biological evolution, they ignore most of the factors that might influence whether a particular individual will reproduce successfully, and instead consider all individuals of a species to be equivalent.
Similarly, EGT models of cultural evolution ignore most of the complexities of human interactions.
For example, rather than reasoning about the physical outcomes of an interaction among several individuals, these
outcomes are represented by payoff values. 
Because the models are highly simplified, they don't give exact numeric predictions of what would happen in real life.

Another problem is that it isn't always easy to get the model right. In EGT research, the authors usually need to be very careful in developing their models, and must convincingly justify their models in their research publications.

On the other hand, a good EGT model can provide explanations of the underlying dynamics of an evolving system, and
establish support for causal relationships. 
Consequently, such models can provide a useful complement to empirical studies, in which there may be questions whether or not a correlation among various factors indicates a causal relationship \cite{aldrich1995correlations}.

In this paper, we begin by providing a tutorial introduction to EGT. We follow this with a description of several cases in which we have used EGT to show how various aspects of a society's structure or its environment can influence the evolution of behaviors and norms within the society. These include: showing that strength of ties and mobility in a society critically affect the evolution of third-party punishment, showing that societal threat directly controls the strength of evolving social norms in a society, and showing that mobility plays a crucial role in the evolution of ethnocentrism.

\section{Background}
\subsection{Evolutionary Game Theory in Biology}
\label{biology}

Evolutionary game theory (EGT) was first developed as an application of game theory to evolving populations composed of multiple animal species, as a way to model how each species' evolutionary fitness causes its proportion of the population to grow or shrink
\cite{smith1973logic}.
The idea is to represent an interaction among animals as a normal-form game. The game's payoffs 
are intended to represent the effect that the interaction will have on the individuals' evolutionary fitness. For example, if two animals fight over a piece of food, one might expect that each individual's fitness would be affected by how the fight affects the animal's health, and whether the animal gets the piece of food.

Rather than developing a detailed model of each specific individual, EGT models typically are at a much more abstract level that does not distinguish among the individuals within each species, but instead looks at the average behavior of all individuals of that species.
More specifically:
\begin{itemize}
\item
If the population is composed of $n$ different species, then for each species $i$ ($i = 1, \ldots n$), all individuals of species $i$ have the same strategy $s_i$, namely the strategy of being a member of species $i$.
This strategy is intended to encompass---in an abstract way, of course---everything that might affect this species' average evolutionary fitness: size, aggressiveness, sensory abilities, intelligence, etc. 
\item
Each species $i$ constitutes some proportion $x_i$ of the entire population, 
with $\sum_{i=1}^n x_i = 1$.
If we choose an individual at random, then for $i=1,\ldots,n$, the probability that this individual uses strategy $s_i$ is $x_i$.
%Thus the entire population can be considered a mixed strategy composed of the relative proportions of the different species.
\end{itemize}
Now, consider an interaction (e.g., a conflict over a source of food) between two individuals: one from species $i$ and one from species $j$.
For simplicity of presentation, we'll restrict this to just two individuals, but it can easily be generalized to interactions among $k$ individuals for arbitrary $k$.

To formulate this interaction as a normal-form game,
let us say that the individuals' expected payoffs are $u(s_i,s_j)$ and $u(s_j,s_i)$,
where ``payoff'' means the effect that the interaction will have on the individual's evolutionary fitness.

The normal-form game is \emph{symmetric}, i.e., if we name the two individuals $a$ and $b$, then it doesn't matter whether the one with strategy $s_i$ is individual $a$ or individual $b$. In either case, this individual's expected payoff is $u(s_i,s_j)$, and the other individual's expected payoff is $u(s_j,s_i)$.

Suppose an individual with strategy $s_i$ meets another individual chosen at random. Then for $j=1,\ldots,n$, the other individual's strategy is $s_j$ with probability $x_j$. Hence the expected payoff for the individual with strategy $s_i$ is
\newcommand{\xvec}{\textbf{x}}
$
\sum_{j=1}^n x_j u(s_i,s_j).
$
Earlier, we said the expected payoff is intended to represent the interaction's effects on evolutionary fitness. The idea is that species $i$'s expected payoff is higher than that of the entire population, then species $i$ will reproduce at a higher rate, hence its proportion $x_i$ will increase. If  species $j$'s expected payoff is lower than that of the entire population, then species $j$ will reproduce at a lower rate, hence $x_j$ will decrease. 

The best-known way to model this is the \emph{replicator dynamic} \cite{taylor1978evolutionary}. The original version is a differential equation that assumes an infinite population and continuous time. 
Let $\pi_i(\textbf{x}) \geq 0$ be the average payoff obtained by individuals of species $i$ when the proportions of each species are
$\xvec = (x_1, \ldots, x_n)$. Then the average payoff for the entire population is
$
\theta(x) = \sum_{i=1}^{n} x_i \pi_i (\xvec).
$
According to the replicator dynamic, the rate of change in each $x_i$
is given by the following differential equation:
\[
dx_i / dt = x_i (\pi_i(\xvec) - \theta (\xvec)).
\]
The replicator dynamic is consistent with the Lotka-Volterra equations for the dynamics of biological systems. Indeed, the replicator dynamic is mathematically equivalent to a generalization of those equations \cite{page2002unifying}.

The replicator dynamic can be translated into a difference equation in which the population is finite, and time proceeds as a sequence of discrete iterations
\cite{hofbauer1984evolutionstheorie}.
This formulation can be used to run a discrete-event computer simulations and look at their outcomes---which is useful
if the differential equations are too complicated to solve mathematically.

One big limitation of the above approach is that it
assumes that the species are \emph{well-mixed}, i.e., that they are uniformly distributed geographically. Such an assumption is often inaccurate; there are many settings in which an individuals' location can make a huge difference in what interactions they have, and how those interactions affect their evolutionary fitness. To model such situations, it often is useful to locate the individuals in a network in which they are restricted to interact with their neighbors. We'll discuss this further in the next section.

\subsection{Modeling Cultural Evolution}
\begin{figure}[t]
\begin{center}
\begin{minipage}{2.1in}
\includegraphics[width=1.8in]{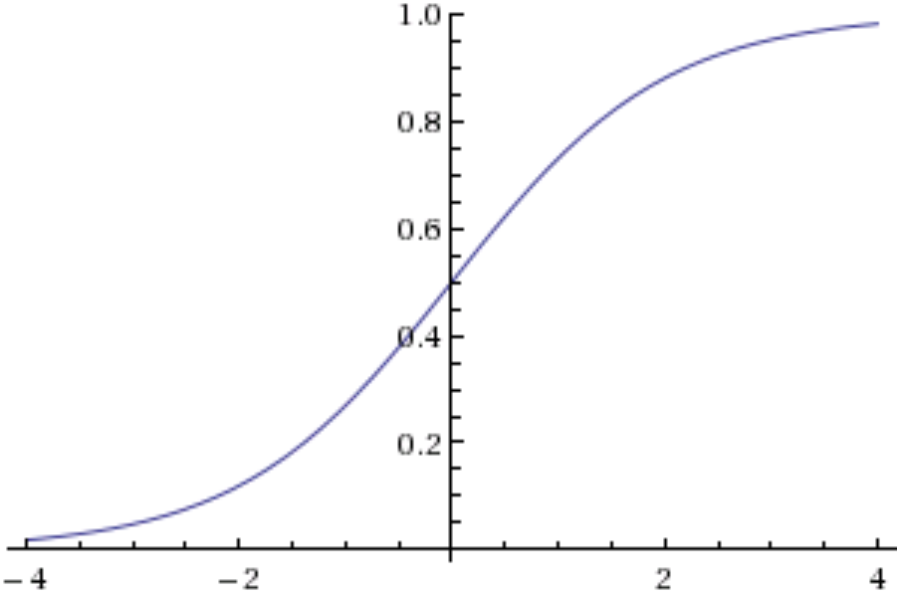}
\caption{Graph of $\frac{1}{1 + e^{s(\pi_i - \pi_j)}}$, for
$s = 1$ and $-4 \leq \pi_i - \pi_j \leq 4$.}
\label{s-curve}
\end{minipage}
%\hspace{\fill}
\hspace{10mm}
\begin{minipage}{.47\textwidth}
\centering
\includegraphics[width=2in]{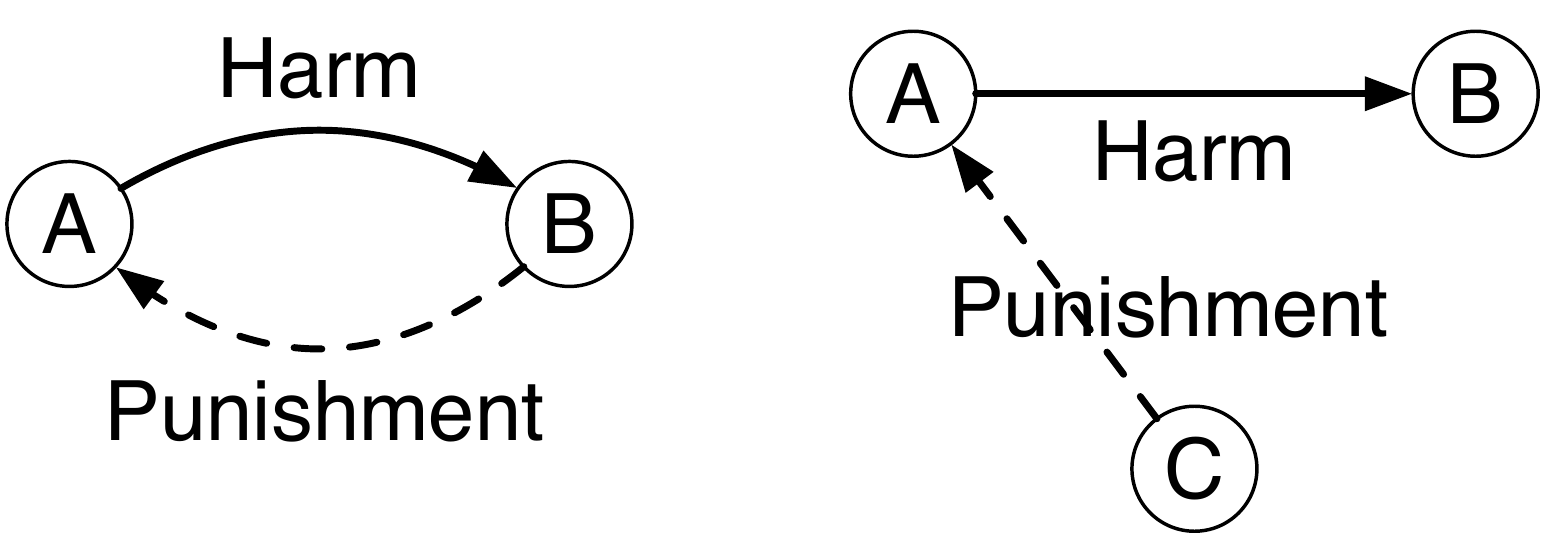}\\[1ex]
(a)\hspace*{.9in}(b)
\caption{(a) direct punishment, and (b) third-party punishment.}
\label{3ppfig}
\end{minipage}
\end{center}
\end{figure}

EGT can be used to model aspects of the evolution of human cultures.
Here, strategies correspond not to collections of individuals, but instead to possible behaviors.
A successful strategy---i.e., a behavior that produces good results---is likely to be adopted by others, hence become more prevalent in the population. Conversely, the prevalence of an unsuccessful strategy is likely to decrease. 
The propagation of these strategies corresponds not to biological reproduction, but instead to \emph{cultural transmission}, in which humans imitate others and learn from others.
Rather than the replicator dynamic, here the evolutionary model is a comparison process,
e.g., a modified version of
the Fermi rule from statistical mechanics \cite{blume1993statistical}.
At each iteration $t$, each individual $i$ uses some strategy in a game-theoretic interaction and receives a payoff $\pi_i$.
Then, before the beginning of iteration $t+1$, $i$ compares $\pi_i$ to the payoff $\pi_j$ received by a randomly chosen neighbor $j$, and decides whether to keep using the same strategy that it used before, or switch to the neighbor's strategy. The probability of switching is given by a version of the well-known sigmoid function (see Figure \ref{s-curve}):
\[
\Pr[i \textrm{ switches to } j\textrm{'s strategy}] = 1/(1 + e^{s(\pi_i - \pi_j)}),
\]
where 
$\pi_i$ and $\pi_j$ are $i$'s and $j$'s payoffs in the current iteration, and
$s \geq  0$ is an arbitrary constant called the \emph{selection strength}.
The Fermi rule can easily be adapted to situations in which the
population isn't well-mixed (see the discussion at the end of Section \ref{biology}).
For example, we can locate the individuals at the nodes of a network, restrict each individual $i$ to interact only with its neighbors, and restrict $i$ to compare its payoff with the payoffs of its neighbors.

Usually the Fermi rule is modified by introducing an \emph{exploration dynamic} that is somewhat analogous to  biological mutation. 
In biological evolution, mutation occurs so rarely that game-theoretic biological models often omit it.
In cultural evolution, an analogous phenomenon happens more frequently: 
individuals to try out new behaviors \cite{traulsen2009exploration}.
The exploration dynamic models this as follows:
when each agent $i$ chooses what strategy to use at the next iteration, there is a small probability $\mu$ that $i$ will choose a strategy $s$ at random from the set of all possible strategies,
regardless of whether $s$ was a successful strategy for the agents who used it in the current iteration,
or whether any agent even used it at all.

\subsection{Tutorial Example}
\label{tutorial}
Many kinds of human interactions have been modeled as social dilemmas in which there is a conflict between behaving responsibly to help maintain a publicly-available resource---e.g., to avoid depleting a renewable resource such as grazing land, fishing stock, clean air, potable water, or climate stability---or to misuse the resource and thereby improve one's own payoff at the expense of others.

The Public Goods Game (PGG)
\cite{henrich2001why,brandt2003punishment,sigmund2010social} provides
an abstract model of the above dilemma.
In this game, $k$ participants are chosen at random from the population, and each is asked to contribute an amount $c$ to a common pool.
There are two pure strategies: \emph{Cooperate} (i.e., contribute) and \emph{Defect} (don't contribute).
The sum of all the contributions is multiplied by a factor $b > 1$
that represents the benefit that the public good provides to the participants,
and the resulting amount is distributed equally to all of the participants, regardless of whether they contributed.

If $m$ of the $k$ participants cooperate, then each cooperator's payoff is $bcm/k - c$. If sufficiently many participants cooperate, all participants will get more than they contributed.
However, as in the famous Prisoner's Dilemma game,\footnote{
\label{pd}
Indeed, when $k=2$ the PGG is equivalent (under standard utility theory) to the Prisoner's Dilemma, i.e., there is a positive affine transformation \cite[Section 1.3]{leyton-brown2008essentials}.}
the payoff for cooperation is lower than the payoff for defection, which is $bcm/k$.
Consequently, when the PGG is played repeatedly with any of the evolutionary dynamics discussed earlier, defection spreads to include nearly 100\% of the population,\footnote{The reason for the word ``nearly'' is that the exploration dynamic keeps the fraction of punishers below $1 - \mu$.} and everyone's payoff is near zero.

What's missing from the above model is a way to \emph{sustain} cooperation. 
One of the most likely candidates is punishment.
There is substantial evidence that humans are eager to punish those who harm them and that punishment can help sustain cooperation 
\cite{fehr2000cooperation,henrich2001why,brandt2003punishment}.
If we allow defectors to be punished by those that they harmed (see Figure \ref{3ppfig}(a)), then the expected payoff for defecting might be smaller than that for cooperating, in which case cooperation will have evolutionary advantage over defection.

To model punishment of defectors, let us say that for each PGG participant $i$ who defected, another randomly chosen participant $j$ will have an opportunity to punish $i$, by subtracting a fixed amount $\rho>0$ from $i$'s payoff. To do this realistically, the model needs to include a cost for performing the punishment \cite{henrich2006costly}: if $j$ chooses to punish, then $j$ will incur a cost $\lambda$ (with $0< \lambda<\rho$).

In this model, each individual's strategy now consists of two choices: whether to cooperate if asked to do so, and whether to punish a defector if asked to do so. Thus there are four pure strategies:
\begin{itemize}
\item CP (Cooperation with Punishment of defectors);
\item CN (Cooperation with Non-punishment of defectors);
\item DP (Defection with Punishment of defectors);
\item DN (Defection with Non-punishment of defectors).
\end{itemize}
Unfortunately, there are several problems with this model. One is that the model of punishment is incomplete because it doesn't include the possibility of punishing cooperators although that phenomenon has been observed in human societies \cite{herrmann2008antisocial}. That problem is relatively easy to fix, but another problem is more serious:
\emph{why do individuals punish at all?}

Since punishing incurs a cost, those who punish will have slightly lower payoffs than those who don't---so over time, the group will evolve to nearly 100\% non-punishers (strategies CN and DN). As the proportion of punishers decreases, the average payoffs to defectors will become higher than the average payoffs to cooperators---so as before, the population will evolve toward nearly 100\% defectors, and each participant's average payoff will be close to zero.

Game-theoretically, this presents a problem. One can argue informally that individuals continue to punish because it helps to maintain cooperative norms, but the model doesn't show how that happens. This was a problem in the EGT literature for some number of years, until it was solved in 2012 by incorporating an observation that sometimes an individual's decision how to behave isn't an absolute one, but instead is \emph{opportunistic}, i.e., swayed by information about whether misbehavior will be punished 
\cite{dos2010evolution,hilbe2010incentives}.
One way to model opportunistic behavior is to have a probability $0\leq \iota\leq 1$ that a participant in the PGG will be informed, just before the contribution phase, of the \emph{punishment reputation} of the individual who will observe them, i.e., whether or not this individual punishes defectors \cite{hilbe2012emergence}. In this case there are four cooperation strategies:

\begin{itemize}
\item
\emph{Cooperate;}
\item
\emph{Defect;}
\item \emph{Opportunistically Cooperate (OC)}, i.e., behave opportunistically if told the punishment reputation, and cooperate otherwise;
\item \emph{Opportunistically Defect (OD)}, i.e., behave opportunistically if told the punishment reputation, and defect otherwise;
\end{itemize}
where ``behave opportunistically'' means to cooperate if one knows one will be punished for defection, and defect otherwise.

During the punishment phase, each individual has an opportunity to observe another individual's behavior, and decide whether to punish them. There are four punishment strategies:
\begin{itemize}
\item
\emph{Responsible} punishment (punish defectors but not cooperators);
\item
\emph{Antisocial} punishment (punish cooperators but not defectors);
\item
\emph{Spiteful} punishment (punish everyone);
\item
\emph{Non-punishment} (punish no one).
\end{itemize}
In this model, the strategy space is much more complicated than it was before.
Each individual's strategy is a combination of one of the cooperation strategies and one of the punishment strategies,
hence there is a total of 16 possible pure strategies.
Evolutionary simulations show that with this model, populations converge toward a high proportion of
cooperation (mostly of the opportunistic kind) and responsible punishment
\cite{hilbe2012emergence}.

\section{Three EGT Studies of Cultural Evolution}
We now describe three published studies of how structural and external conditions affect the emergence of social norms
\cite{roos2014high,roos2015societal,de2015inevitability}. We focus on explaining the design decisions: how we chose to model things, and why.

\subsection{Evolution of Third-Party Punishment}
\label{3pp}

\paragraph{Motivation.}
In Section \ref{tutorial} we discussed the question of why direct punishment should exist, since it incurs a cost to the punisher. 
In \emph{third-party punishment} (3PP),  an individual \textsf{A} harms another individual \textsf{B}, and
an uninvolved third party \textsf{C} punishes \textsf{A} for causing that harm (see Figure \ref{3ppfig}(b)).
There is significant empirical evidence that humans \cite{dequervain2004,bernhard2006parochial} and non-human species
\cite{raihani2010punishers}
 are willing to engage in 3PP,
but this is even more of a puzzle than direct punishment: not only does it incur a cost to the punisher, but the punisher was not involved in the original interaction.

%3PP can arguably be more effective at norm maintenance than direct punishment, since a norm-violator might be punished by multiple other agents in a population of third-party punishers, and 
 
%In Section \ref{tutorial} we discussed the question why direct punishment should exist, since it incurs a cost to the punisher. 3PP is even more of a puzzle: not only does it incur a cost, but the punisher was not harmed by the original interaction.

Since the work described in Section \ref{tutorial} showed that punishment reputation was a key ingredient in the evolution of responsible direct punishment, we decided to investigate whether punishment reputation could also account for the evolution of responsible 3PP. 
Below we summarize the results of that investigation (see \cite{roos2014high} for details).

We began with the model in Section \ref{tutorial}, and modified it as follows:
\begin{itemize}
\item
For the contribution phase, we changed the game to a cooperation dilemma (the special case $k=2$, as discussed in Footnote \ref{pd}) between two individuals.
\item
For each individual in the cooperation dilemma, we allowed an uninvolved third party observe that individual's behavior and decide whether to punish them---hence 3PP rather than direct punishment
\item
The four punishment strategies were the same as in Section \ref{tutorial}, except that the punishment was 3PP as described above.
\item
The opportunistic strategies look at the punishment reputations of the individual's neighbors,
and decide whether to cooperate by looking at the expected payoffs for cooperating and defecting.
\item
The punishment reputations of the neighbors are always available, making the OC and OD strategies identical. Hence there are just three pure strategies for contribution: Cooperation, Defection, and Opportunism.
\end{itemize}
%What we found was that with those modifications, 3PP did not evolve. Instead, it nearly died out (again, the reason it didn't fully die out was because of the exploration dynamic).

Next, drawing on classic sociological and psychological theory, we hypothesized that social-structural constraints in human populations might play a crucial role in enabling the evolution of responsible 3PP. 
The constraints that particularly interested us were strength-of-ties  and residential mobility, which have wide-ranging consequences for humans
\cite{granovetter1983strength,oishi2007residential,oishi2010psychology}.

We hypothesized that punishment reputation was more likely to
incentivize cooperation and responsible 3PP in situations where there was high strength-of-ties (individuals interacted primarily with a small group of other individuals) and low residential mobility (no individual would be able to exit the group easily).

As a basis for modeling strength-of-ties and residential mobility, we began by
locating the individuals as nodes on a network (see the discussion at the end of Section \ref{biology}).
We used
Watts-Strogatz small-world networks \cite{watts1998collective}.
On this network, we implemented strength-of-ties and residential mobility as follows:
\begin{itemize}
\item
{\em Strength-of-ties.} 
Granovetter \cite{granovetter1983strength} measured tie strength between two humans in terms of how often they interacted with each other during a period of time. In our EGT model, each individual generally has an equal number of interactions at each iteration, so an individual with few connections has a relatively high number of interactions with those neighbors (hence high strength-of-ties), while an individual with many connections has a relatively low number of interactions with a greater variety of individuals (hence low strength-of-ties).
Thus if an individual is located at a node with degree $d$, then the strength-of-ties to each of the individual's neighbors is $1/d$.
\item
{\em Residential mobility.}
Residential mobility \cite{oishi2007residential} is the degree to which humans are able to change their location, and, as a result, their position within the social network within a population. Some human populations, particularly those that are individualistic, have very high mobility where people can easily exit the group, whereas others, particularly collectivistic cultures, are much more dependent on others and are less able to easily exit the group
\cite{oishi2010psychology,schug2009similarity,schug2010relational}. We implemented a simple model of residential mobility using a probability $m$ with which, at the beginning of each generation, an individual would switch position with a randomly chosen other individual in the population.
\end{itemize}

\begin{figure}
\includegraphics[width=0.48\textwidth]{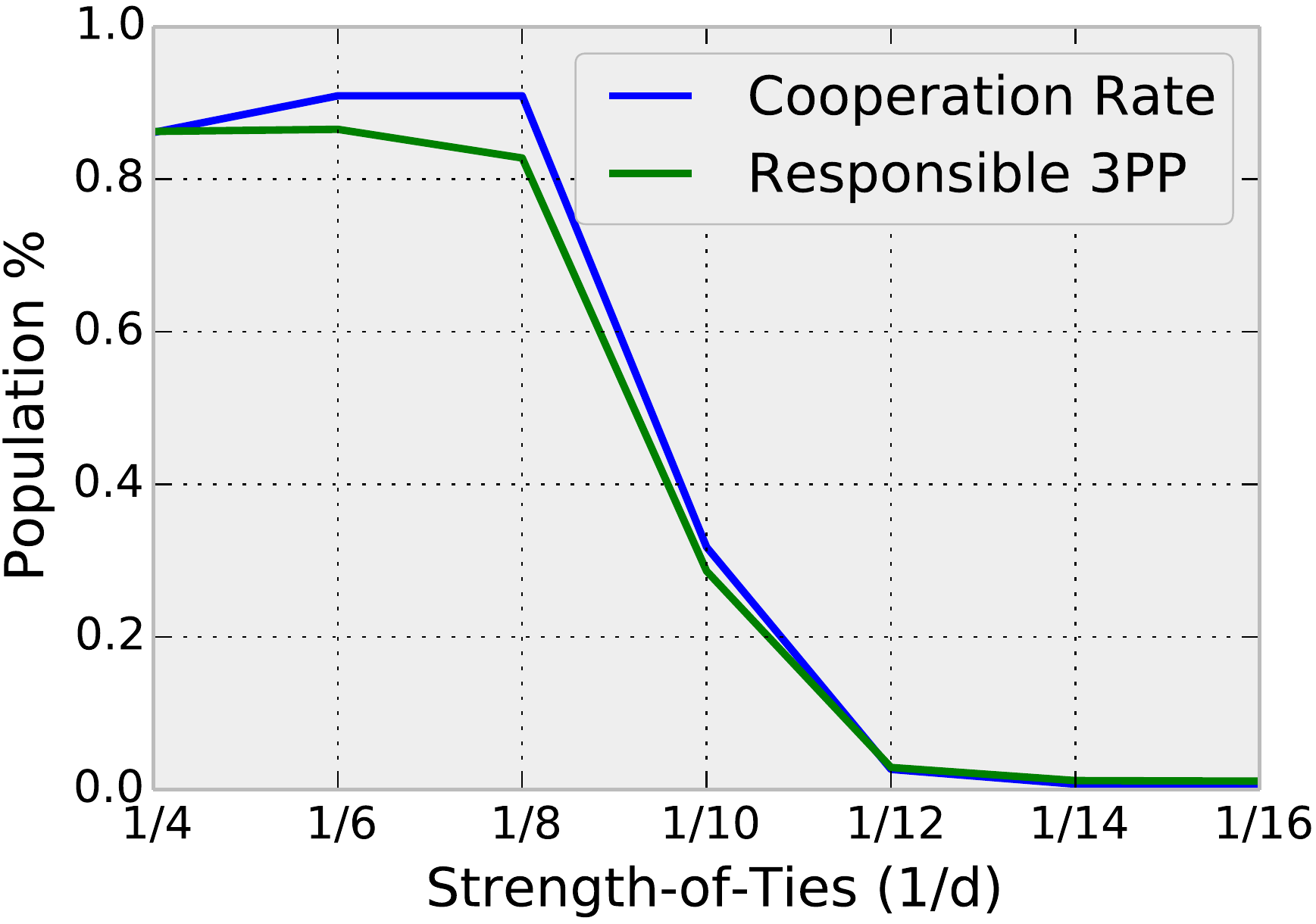}\hspace{\fill}\includegraphics[width=0.48\textwidth]{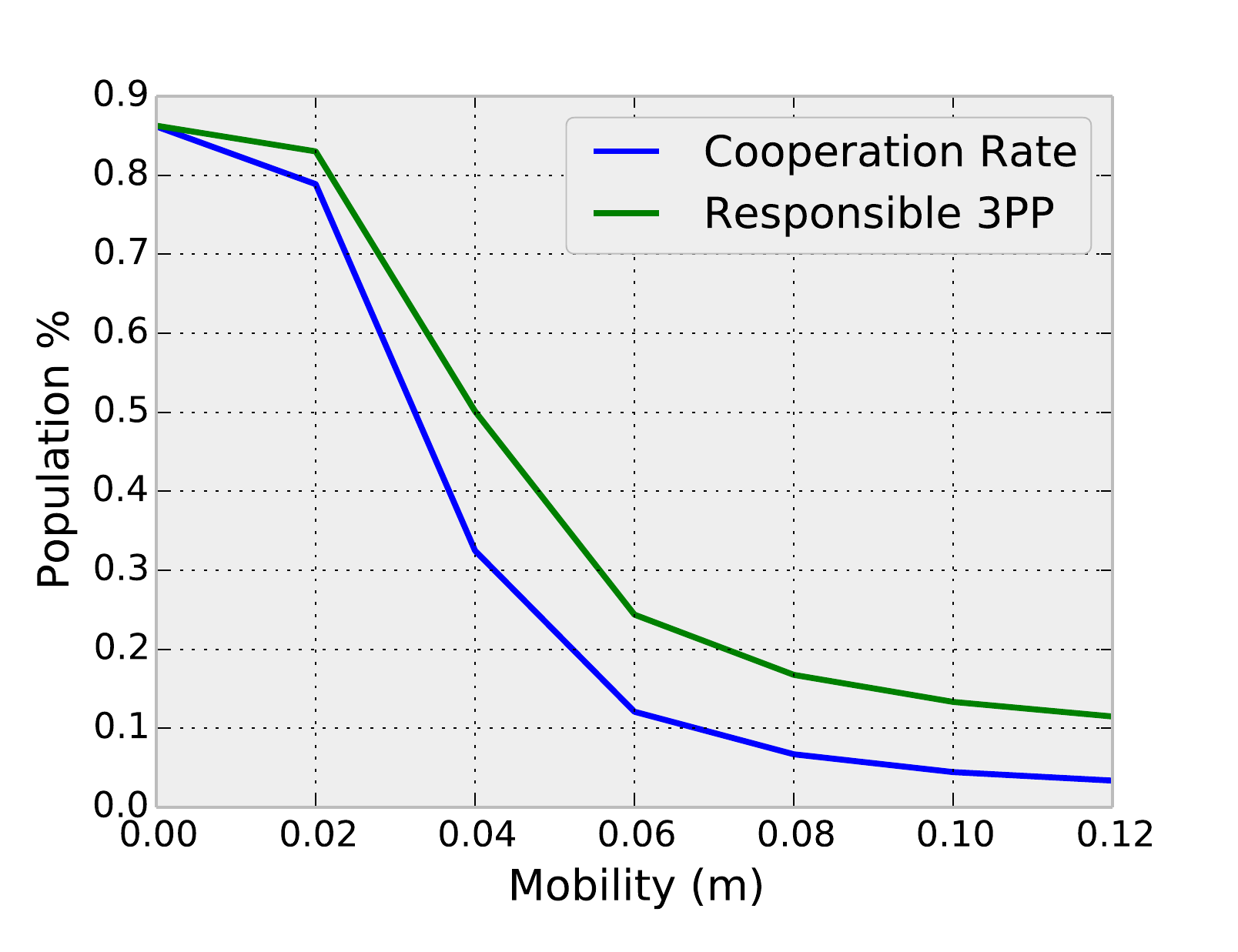}
\caption{Effects of strength-of-ties (a) and mobility (b) on the proportions of cooperators and responsible 3rd-party punishers.
Long run average proportions were attained by averaging 100 simulation runs over 5000 generations for populations of 1000 individuals, with cooperation contribution $c=1$, cost $\lambda = 1$  to punish, punishment penalty $\rho = 3$, exploration rate $\mu = 0.01$, selection strength $s = 0.5$.
In (a) the mobility is $m = 0$, and in (b) the strength-of-ties is 1/4.
For additional details see \cite{roos2015societal}.}
\label{3ppresults}
\end{figure}

We ran a large number of EGT simulations in which we varied the individuals' average strength-of-ties and their average mobility.
The simulation results confirmed our hypothesis: the evolution of responsible 3PP depended critically on conditions of high social-structural constraint, i.e. high average strength-of-ties and low mobility (see Figure \ref{3ppresults}).

Examination of the results showed that when responsible 3PP evolves, it does so as an ultimately non-altruistic trait. The behavior acts as a signal to potential co-players in the neighborhood that non-cooperation will not be tolerated. High strength-of-ties and low mobility allow clustered agents engaging in responsible 3PP to induce cooperation in their neighborhood. By inducing such local cooperation, clusters of 3PP agents increase their own payoff and spread. This leads to the emergence of responsible 3PP in the population as a whole. By contrast, low strength-of-ties and high mobility prevent 3PP agents from inducing a local culture of cooperation, and hence responsible 3PP does not
evolve. 

\subsection{Societal Threat and the Strength of Social Norms}
\label{obhdp}

The strength of social norms varies widely around the globe. For example there is considerable cross-cultural variation in norms for fairness, cooperation, and the willingness to punish to enforce such norms 
\cite{balliet2013trust,gelfand2011differences,henrich2006costly,herrmann2008antisocial};
and there is evidence that these variations are related to the degree of threat to the society---which can include manmade threats such as threats of invasions,
low natural resources (e.g., food supply),
and high degrees of natural disasters such as floods, cyclones, and droughts
\cite{cosner1956functions,lomax1972evolutionary,gelfand2011differences}.
What has been unclear is whether the degree of threat might actually require stronger norms and associated punishment of deviance in order to survive under high threat---or more generally, whether differences in punishment across societies have any evolutionary basis.

We now describe a study in which we used
EGT modeling to explore whether higher levels of threat cause stronger norms of cooperation, with higher punishment of deviance, to be evolutionarily adaptive (see \cite{roos2015societal} for details).
We studied both cooperative norms (as in Section \ref{tutorial}) and coordination norms (where individuals need to coordinate their actions in order to accomplish something). Below we discuss only the former; for the latter, see \cite{roos2015societal}.

For cooperative norms, our EGT model was similar to one in Section \ref{tutorial}. Individuals interacted in a PGG with a contribution phase and a punishment phase.
However, we made the following modifications:
\begin{itemize}
\item
Our previous study (the one in Section \ref{3pp}) having convinced us of the importance of 
geographically constraining the individuals on a network, we also did that here.
However, we didn't need to vary degree of the nodes this time, so we used a simple grid, which is the kind of network that's used the most frequently in EGT studies \cite{hammond2006evolution,hartshorn2013evolutionary}.
\item
The contribution strategies were the same as in Section \ref{3pp}:
Cooperation, Defection, and Opportunism, where the latter strategy looks at the punishment reputations of the individual's neighbors,
and decides whether to cooperate by looking at the expected payoffs for cooperating and defecting.
\end{itemize}
As a measure of the norm strength in a population, we used the percentage of agents who adhere to the norm, and the proportion of punishers.

% Our results are comprised of the evolutionary outcomes of the process of cultural adaptation, including the evolutionary trajectories of behaviors and the composition of behaviors in populations that have arrived at a stable state. We determine these results through multi-agent computer simulations.

%\emph{(explain motivation for next two paragraphs: need generic notion of "threat", and the economic "diminishing returns" was an intuitive fit.)}

To implement the notion of societal threat, we started by giving
all individuals a base payoff from the environment, to capture the notion that our game models just one of the many kinds of interactions that individuals might have in a more realistic world. Then we subtracted an amount $\tau$ from this base payoff, to model the level of threat. Since threats like drought, hurricanes, tornadoes, famine, or hostile invasions all can reasonably be assumed to reduce the general payoff that agents in a population receive from their environment, this captures the essential effect of a broad array of threats. For instance, ecological threats are related to the availability of natural resources in that they often diminish agricultural yields and engender food shortages \cite{popp2006effects}, and managing them often requires the use of the population's resources.

After each individual receives its base payoff and the payoffs from the contribution and punishment phases, the total acquired payoff $p$ is transformed to reproductive fitness through a fitness function
$f(p) = 1 - e^{-0.1p}$,
which is intended to capture the principle of diminishing marginal utility: increased payoffs produce diminishing marginal increases in fitness, e.g., the fifth meal of a day is not as important to an agent's fitness or wellbeing as the first. There is an abundance of evidence across disciplines that increased resources lead to diminishing increases in utility \cite{diener2002will,diener2010income,foster2004diminishing}, and diminishing-returns fitness functions have been used in several other EGT studies \cite{foster2004diminishing,godfray1991signalling,grodzinski2011parents}.

\begin{figure}[t!]
\centering
\includegraphics[width=.8\textwidth]{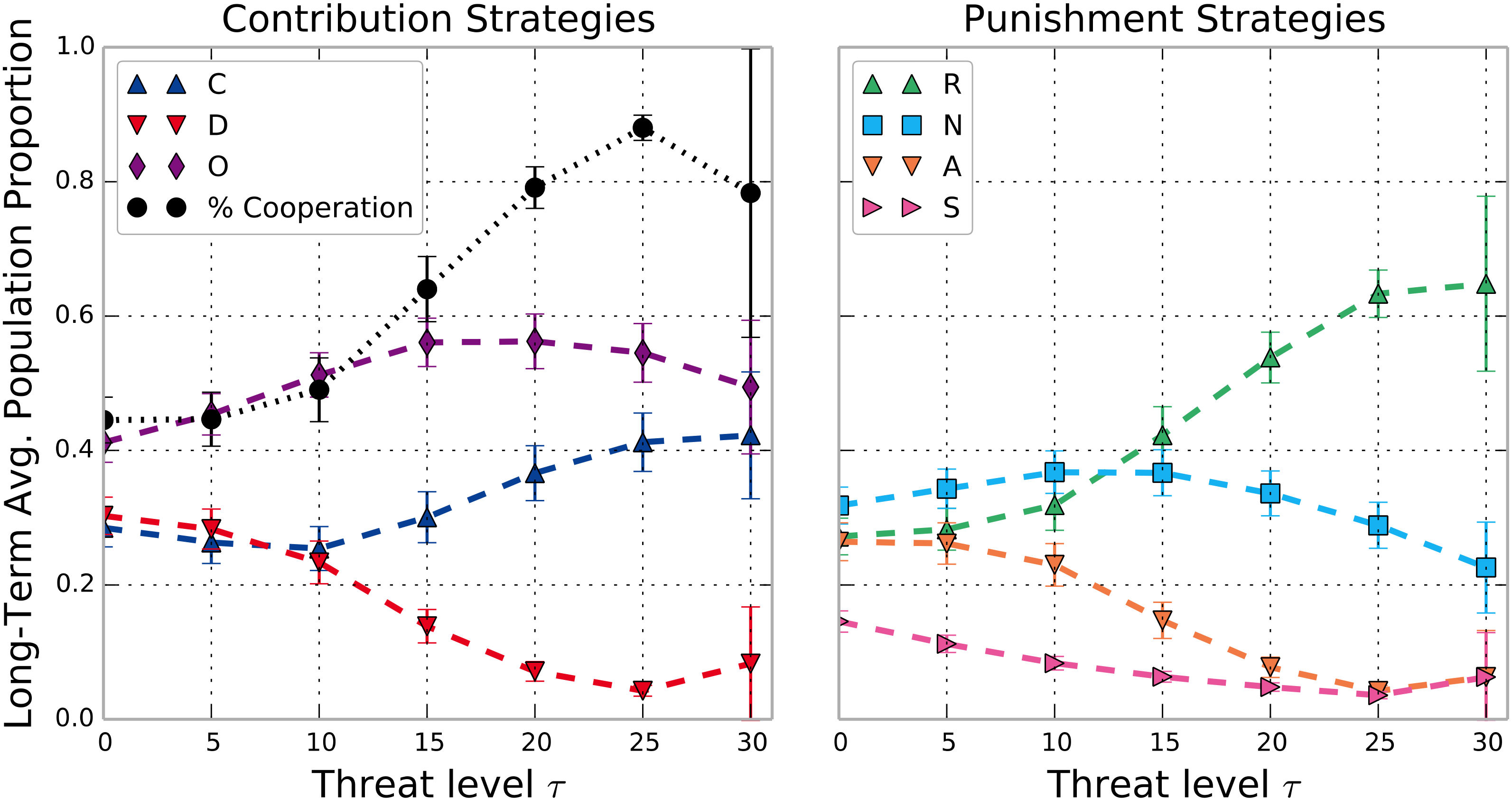}
\caption{Long-term average proportions of cooperation (at left) and punishment (at right) strategies under different levels of threat. The graph at left also includes the cooperation rate, the percentage of actions in the game that are cooperation. To attain long-term average proportions, we ran simulations for 5000 timesteps and averaged the population proportions over all timesteps. Each data point in the plot is the average of 50 simulation runs. Error bars show standard deviation. Model parameters are $r = 3$, $c = 1$, $\lambda = 1/2$, $\rho = 3/2$, $\mu = 0.1$, $d = 0.1$, base-pay = 30. Note higher Responsible Punishers (R) and higher cooperation by Opportunistic (O) and Cooperator (C) agents under higher threat levels.
For additional details see 
\cite{roos2015societal}.}
\label{obhdpfig}
\end{figure}

We ran EGT simulations with the above model to test the hypothesis that exposure to threat is an important causal factor in the emergence of strong norms under the evolutionary pressures of cultural adaptation. If exposure to threat is indeed a cause of the emergence of
strong norms, agents that adhere to the norm and enforce the norm through punishment should thrive evolutionarily under higher threat, while other agents that do not enforce the norm should not fare well evolutionarily. In contrast, under lower threat, we would expect the evolutionary pressures to allow for a greater variety of strategies and less norm-enforcing punishers.
The results (see Figure \ref{obhdpfig}) confirm this hypothesis.

\subsection{Why Has Ethnocentrism Decreased?}
\label{ethnocentrism}

Almost all current and historical conflicts around the world can be characterized by individuals defining themselves and others in terms of their group membership. This is supported by substantial empirical evidence that show that individuals have a natural tendency towards ethnocentrism, i.e., to favor in-group members, while being hostile towards out-group individuals \cite{brewer1985psychology,chen2009group,hewstone2002intergroup}. Several EGT studies have also shown that group-biased behavior among individuals emerges readily and dominantly in populations comprised of various groups \cite{antal2009evolution,fu2012evolution,hammond2006evolution}. 

Although these studies might seem to indicate that ethnocentric behavior among humans is inevitable, statistics show that over the past few centuries of human civilization, violence and out-group conflict have declined dramatically. It occurred to us that the seeming disconnect between these statistics and the research literature might be resolved by examining how residential mobility (see Section \ref{3pp}) affects the evolution of ethnocentrism. 
Below we summarize our EGT study on this topic; the details are described in \cite{de2015inevitability}.

Classic sociological and psychological theory suggests that in high-mobility contexts, individuals form new relationships and sever unwanted ones with great ease, and are more open towards strangers---but in low-mobility contexts, individuals have fewer opportunities to form new relationships, and severing existing relationships can lead to being ostracized from one's only social circle \cite{oishi2015psychology,landrine1992clinical}. 
Thus we hypothesized that residential mobility might counteract the evolution of ethnocentrism, by discouraging \emph{group-entitative} behavior (in which one's behavior toward others depends primarily on what groups they are in) and encouraging  \emph{individual-entitative} behavior (in which one reacts to others primarily as individuals rather than as members of a particular group).

For our model, we began with the one in Hammond and Axelrod's pioneering EGT study of ethnocentrism \cite{hammond2006evolution}.
In their model, agents were located on a grid, and interacted with their immediate neighbors.
Each agent had a perceivable \emph{group tag} to identify what group it was in, and each agent $i$ could behave differently toward \emph{in-group members} (members of $i$'s group) and \emph{out-group members}(members of other groups).
At each iteration, each agent had a Prisoner's Dilemma interaction with each of its immediate neighbors. The possible strategies were CC (cooperate with both in-group and out-group members), CD (cooperate with in-group and defect with out-group),  DC (defect with in-group and cooperate with out-group), and DD (defect with both in-group and and out-group). In their study, they tested a wide range of parameter settings, and in each case ethnocentrism (modeled by the CD strategy) became predominant.

To test our hypothesis, we needed to modify Hammond and Axelrod's model to introduce mobility. 
Furthermore, in their model, all of an agent's interactions were group-entitative, and we needed to introduce the possibility of individual-entitative behavior.
To accomplish this, we made the following modifications:
\begin{itemize}
\item
We implemented mobility as in Section \ref{3pp}: we used a probability $m$ with which, at the beginning of each generation, an individual would switch position with a randomly chosen other individual in the population. 
\item Each individual could either be \emph{group-entitative}, i.e. base their actions solely on what the group an individual is in, or \emph{individual-entitative}, i.e., base their actions on knowledge of each individual rather than their group tag.
\item The action of a \emph{group-entitative} individual $i$ towards an individual $j$ depended only on $i$'s last encounter with anyone from $j$'s group. Thus, $i$ had two possibly different strategies: one for in-groups and another for out-groups. Each of those strategies was one of the following: \emph{AllC} (always cooperate), \emph{AllD} (always defect), \emph{TFT} (Tit-for-Tat: play whatever action the opponent played in $i$'s last interaction with anyone from $j$'s group), or \emph{OTFT} (play the opposite of what \emph{TFT} would play).
\item The action of an \emph{individual-entitative} agent $i$ depended only on its last encounter specifically with agent $j$, ignoring $j$'s group tag. Thus, $i$ could have one of the above four strategies, except that \emph{TFT} and \emph{OTFT} depended on $i$'s last interaction with $j$ specifically, rather than someone in $j$'s group.
%\item We deleted the punishment phase used in Sections \ref{3pp} and \ref{obhdp}. 
\end{itemize}

\begin{figure}[t!]
\includegraphics[width=0.49\textwidth]{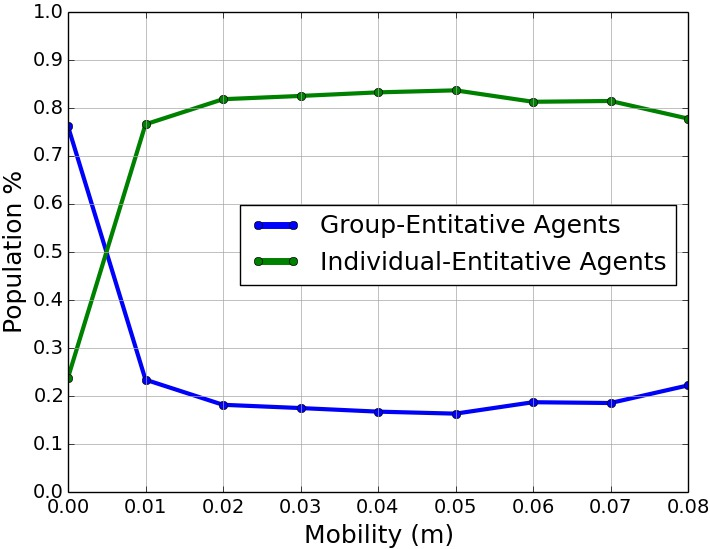}\hspace{\fill}\includegraphics[width=0.49\textwidth]{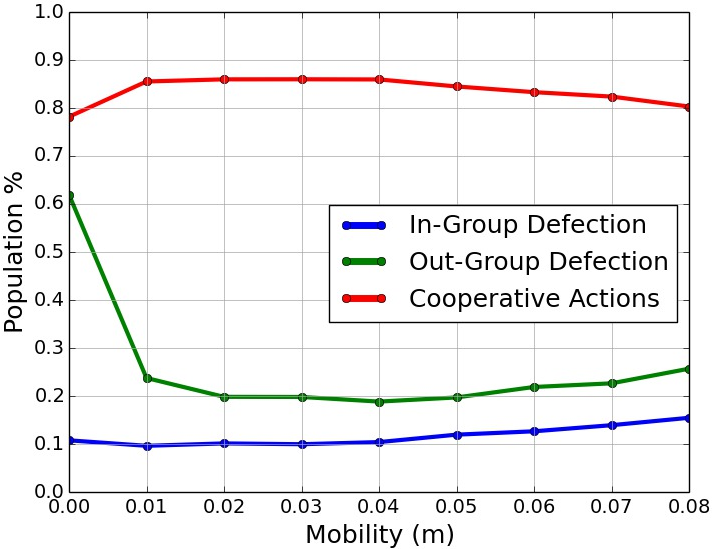}\\
\mbox{}\hspace{\fill}(a)\hspace{\fill}\hspace{\fill}(b)\hspace{\fill}\mbox{}
\caption{Proportions of agents and actions as a function of mobility, after 30000 iterations.  Each
data point is an average of 100 simulation runs. The plots show the proportions of (a) the group-entitative and individual-entitative agents, and (b) the actions played by the agents. For additional details see
\cite{de2015inevitability}.}%, (c) the strategies of the individual-entitative agents, and (d) the in-group and (e) out-group strategies of the group-entitative agents.}
\label{ethnoresults}
\end{figure}

The results of our EGT simulations (see Figure \ref{ethnoresults}) confirmed our hypothesis. With low mobility (low values of $m$), groups tended to cluster together heavily, i.e., agents interacted primarily with in-group members. Thus, group-entitative strategies (in-group cooperation and out-group hostility) were profitable in terms of payoffs and ethnocentric behavior became evolutionarily dominant. However, as we increased $m$, ethnocentrism decreased. Individuals were more likely to interact with out-group members, and hence couldn't rely on high payoffs from in-group interactions. Thus, individual-entitative strategies became evolutionarily dominant with strategies like \emph{TFT} gaining prominence.
% Thus, the results from our model (see Figure \ref{ethnoresults}) confirmed our hypothesis that mobility can counteract showed that ethnocentrism is not inevitable after all.

\section{Discussion}
In this paper, we have discussed how evolutionary game theory can be used as a framework for studying the evolution of cultural norms, and have described the modeling decisions that we made in three EGT studies of the evolution of cultural norms. 
These studies have helped to establish support for causal relationships between several structural factors---strength-of-ties, residential mobility, and societal threat---and societal traits such as third-party punishment, norm enforcement, and ethnocentrism. 

At this point, some readers might be pondering the following question: \emph{why did we use EGT for these studies, rather than conventional game theory?}

Conventional game theory is good for analyzing situations where we know the individuals' preferences, and want to predict what they will do based on those preferences. But in our work, we wanted to know how these preferences arose. We were interested in the following kinds of questions:
\begin{itemize}
\item
What kinds of structural and external factors might have led to the emergence of behaviors we see among individuals in a society?
\item
What evolutionary pressures might have led to variations in those  behaviors?
\item
Can they be validated by observed phenomena?
\end{itemize}
Conventional game theory wouldn't have answered those questions.
To lay out an individual's preferences in a conventional game-theoretic model would, in essence, be building into the model the
very traits whose emergence we were wondering about. We instead needed to lay out the structural/environmental factors that might be responsible for the evolution of those traits, to see whether those traits would evolve.

These results also illustrate how an interdisciplinary team---in this case, computer scientists and psychologists---can achieve results that couldn't have been done without cooperation among participants from both disciplines.
The work
required a strong understanding of both psychological principles and computational and mathematical modeling techniques.

\bibliographystyle{plain}
\bibliography{00refs}
\end{document}